\documentclass[reprint,pra,longbibliography,amsmath,amssymb,nofootinbib,showpacs,twocolumn]{revtex4-1}

\usepackage{graphicx}% Include figure files
\usepackage{epsfig}% Include figure files
\usepackage{amssymb,amsmath,amsfonts}
\usepackage[hidelinks]{hyperref}
\usepackage{overpic} 
\usepackage{wasysym}
\usepackage{latexsym}
\usepackage{array,multirow}
\usepackage{verbatim}
\usepackage[normalem]{ulem}
\usepackage{color}
\definecolor{orange}{rgb}{1,0.45,0}

\newcommand{\dd}{\partial}
\newcommand{\rd}{\mathrm{d}}
\newcommand{\pd}[2]{\frac{\partial #1}{\partial #2}}

\newcommand{\eps}{\epsilon}
\newcommand{\ps}[2]{\langle #1 | #2\rangle}

\newcommand{\beq}{\begin{equation}}
\newcommand{\eeq}{\end{equation}}

\newcommand{\ii}{\mathrm{i}}

\begin{document}
\title{Multiple critical coupling and sensing in a microresonator-waveguide system}%{Multiple critical coupling, resonance depth, and sensing with a microresonator-waveguide system}% extinction ratio

\author{Nirmalendu Acharyya and Gregory Kozyreff\\
{\small Optique Nonlin\'eaire Th\'eorique, Universit\'e libre de Bruxelles (U.L.B.), CP 231, Campus de la Plaine, 1050 Bruxelles, Belgium}}

%\affil[*]{Corresponding author: nacharyy@ulb.ac.be}
%
%\dates{Compiled \today}
%
%\ociscodes{(230.5750) Resonators; (230.7370) Waveguides; (130.6010) Sensors; (130.2790) Guided waves; (130.7408)   Wavelength filtering devices.}
%
%\doi{\url{http://dx.doi.org/10.1364/ao.XX.XXXXXX}}

\begin{abstract}
We study the optical transmission of a waveguide that is side-coupled to a high-$Q$ circular microresonator. The coupling is critical if the intrinsic resonator losses equate the coupling losses to the waveguide. When this happens, the transmittance of the waveguide displays resonance dips with maximal depth as the frequency is swept through the resonators resonances. We show that multiple configurations, parameterised by the minimal distance between the resonator and the waveguide, can lead to critical coupling. Indeed, for a sufficiently large resonator radius, the flow of power between the waveguide and the resonator can change sign several times within a single pass. This leads to an oscillatory coupling parameter as a function of the separation distance. As a result, multiple geometrical configurations can lead to critical coupling, even if the waveguide lies in the equatorial plane of the resonator. These results are explained using coupled-mode theory and full wave numerical simulations. In the vicinity of secondary or higher-order critical coupling, the depth of the transmittance dip is very sensitive to the environment. We discuss how this effect can be exploited for sensing purpose. Alternatively, by actively controlling the environment in the secondary critical configuration, the waveguide/resonator system can be driven as an optical switch.
\end{abstract}

%\setboolean{displaycopyright}{true}

\maketitle
%\thispagestyle{fancy}
%
%\ifthenelse{\boolean{shortarticle}}{\ifthenelse{\boolean{singlecolumn}}{\abscontentformatted}{\abscontent}}{}
%

%%%%%%%%%%%%%%%%%%%%%%%%%%%%%%%%%%%%%%%%%%%%%%
%%%%%%%%%%%%%%%%%%%%%%%%%%%%%%%%%%%%%%%%%%%%%%
\section{Introduction}
Circular resonators such as micro-spheres, micro-toroids, wedge-resonators and micro-rings have dramatically improved the quality of light-matter interaction in cavities, in the sense of enhanced interaction strength and spectral purity of the recorded signals. With losses only limited by intrinsic material absorption, quality factors $Q$ reaching $3 \times10^{11}$  have been demonstrated~\cite{Savchenkov-2007b}. The detection of small wavelength shifts of the resonances of these cavities is the basis of very sensitive detectors~\cite{Arnold-2003,Vollmer-2008,He-2011,Vollmer-2012,Frenkel-2016}. Nonlinear effects are also enhanced, with reduced threshold for parametric oscillations~\cite{Kippenberg-2004,Savchenkov-2004b,Furst-2010b} and second harmonic generation~\cite{Ilchenko-2004} and with a strong current focus on frequency comb generation~\cite{Chembo-2010,Okawachi-2011}. In this respect, at the photon level, phase matching corresponds to angular momentum conservation and follows the associated composition rules of quantum mechanics~\cite{Kozyreff-2008}. Thanks to the high $Q$, surface second harmonic generation mediated by as few as one hundred \emph{small} molecules (equivalent in mass to a single protein) has been demonstrated~\cite{Dominguez-2011}. Furthermore, the large $Q/V$ ratios, where $V$ is the mode volume, makes these cavities particularly useful to study quantum electrodynamics~\cite{Aoki-2006} and quantum optics~\cite{Strekalov-2016}. Currently, the field is steadily progressing towards integrated application, with high-$Q$ micro-resonators being demonstrated with silicon~\cite{Bogaerts-2012,Soltani-2016} and silicon nitride platforms~\cite{Tien-2011,Spencer-2014,Xuan-2016,Ji-2017}.

The most effective way to  pump and interrogate the resonances of these resonators is to couple them with a waveguide. A detailed theory of this coupling has been worked out and demonstrated experimentally before \cite{Gorodetsky-1999,Cai-2000,Yariv-2000}, with special emphasis on critical coupling, where ideally all the optical power injected in the waveguide can  be dissipated by the resonator. The external waveguide, through its coupling to the resonator, also represents an adjustable loss mechanism and can therefore be a key parameter in the hands of an experimentalist, notably to control the threshold of optical parametric oscillations and quantum light production~\cite{Furst-2011}.

\begin{figure}
\includegraphics[width=8.5cm]{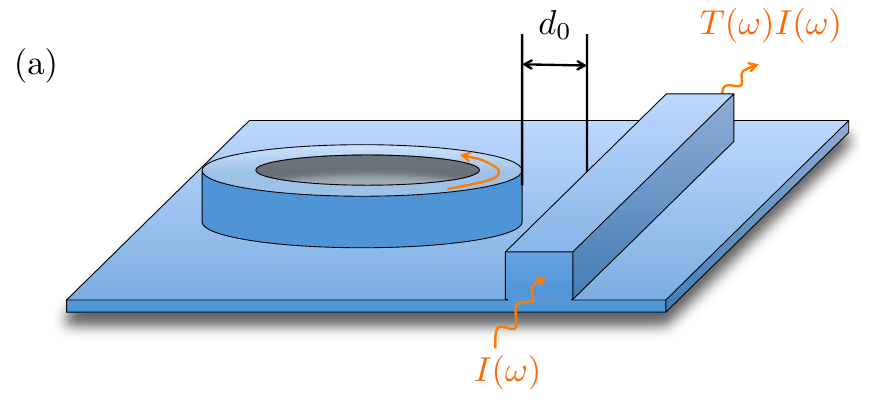}\\
\includegraphics[width=8.5cm]{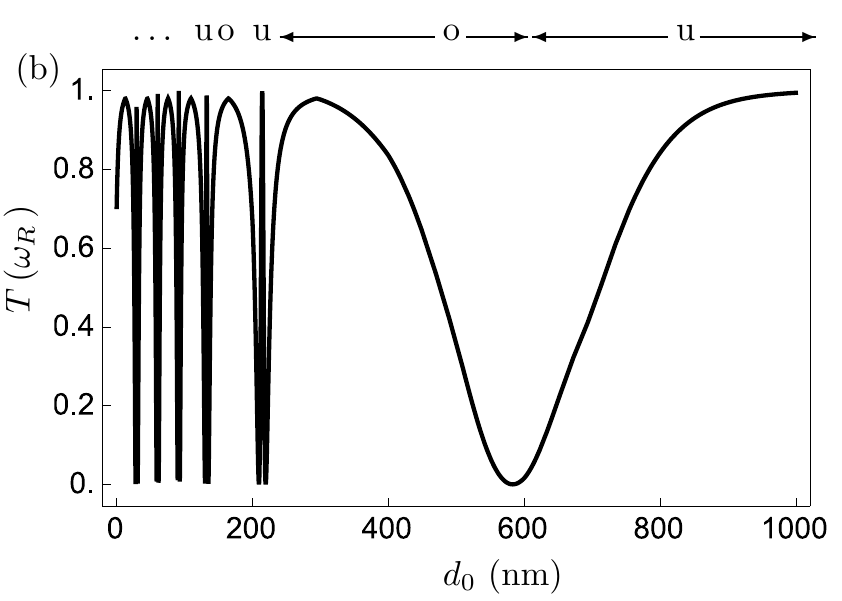}
\caption{(a) Schematic of the waveguide-resonator system. (b) Transmittance at resonance ($\omega=\omega_R$) as a function of coupling distance $d_0$ for a 100\,$\mu$m-radius silicon (refractive index=3.48) ring coupled to a ridge waveguide  on a SiO$_2$ substrate (refractive index=1.45). Microring and waveguide width: $w=200$\,nm. Height: 400\,nm. Base height: 50\,nm.  $Q=4.4\times 10^5$, $\lambda\approx1.55\mu m$, $|\tilde\alpha|=0.99$. The near-zero transmittance at $d_0\approx600$\,nm is the usual critical coupling situation. Additional critical coupling distances are found for $d_0<210$\,nm, always appearing in pairs. The letters `u' and `o' indicate under- and over-coupled region, respectively.}
\label{fig:schematic1}
\end{figure}

Recently, it has been pointed out that multiple critical configurations can exist if the waveguide is buried under the micro-resonator~\cite{Ghulinyan-2013,Turri-2016}. As the vertical gap between the waveguide is decreased, a pair of critical coupling configurations are demonstrated, in addition to the usual one. Moreover, within the narrow range of gaps between the two newly found critical coupling distances, the system reverts to an under-coupling state. Until now, it was assumed that such an exotic situation was only possible if the waveguide lies in a different plane from that of the resonator~\cite{Ghulinyan-2013,Turri-2016}. However, this is not the case: we show that similar multiple critical coupling can be found with micro-resonators \textit{side coupled} to a waveguide. Moreover, our analysis shows that an arbitrary number of critical coupling configurations can be achieved, depending on the micro-resonator radius, see Fig.~\ref{fig:schematic1}.  We demonstrate this possibility using coupled-mode theory and finite element numerical simulations, which are in full quantitative agreement.

As can be seen in Fig.~\ref{fig:schematic1}, the transmission at resonance, $T(\omega_R)$, exhibits very sharp features in the multi-critical regime when plotted as a function of the waveguide-resonator distance. In this paper, we will show how this can be exploited for sensing purpose. Indeed, the   transmission increases very rapidly as soon as one departs from the conditions of critical coupling. Rather than to detect the shift of spectral resonance as a response of environmental change, we propose to detect the increase of transmission, or changes in extinction ratios. 

In what follows, we first briefly review the general derivation of the intensity transmittance of the waveguide/microresonator system. This will be necessary to explain and derive simple analytical estimates of the emergence of multiple critical coupling. Next, we discuss the exploitation of this effect for sensing or switching purposes. Finally, we conclude.

%------------------------------------------
\section{Theory of Multiple Critical Coupling}

\begin{figure}[h!]\centering
\includegraphics[width=8.cm]{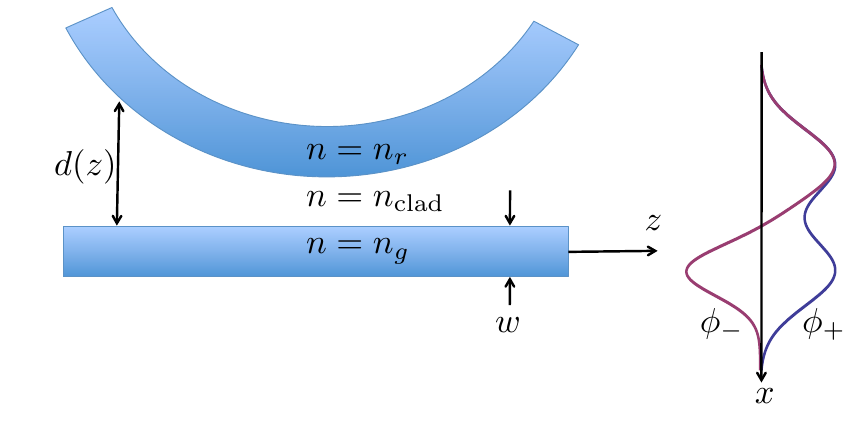}
\caption{Schematic of the coupling region and the local profiles of the propagation eigenmodes $\phi_{\pm}$. between a waveguide and micro-resonator. $n_g$: guiding refractive index; $n_\text{clad}$: cladding refractive index; $n_r$: resonator refracting index. Throughout this paper, we will assume that $n_r=n_g$}
       \label{fig:topview}
  \end{figure}

\begin{figure}
\includegraphics[width=7.5cm]{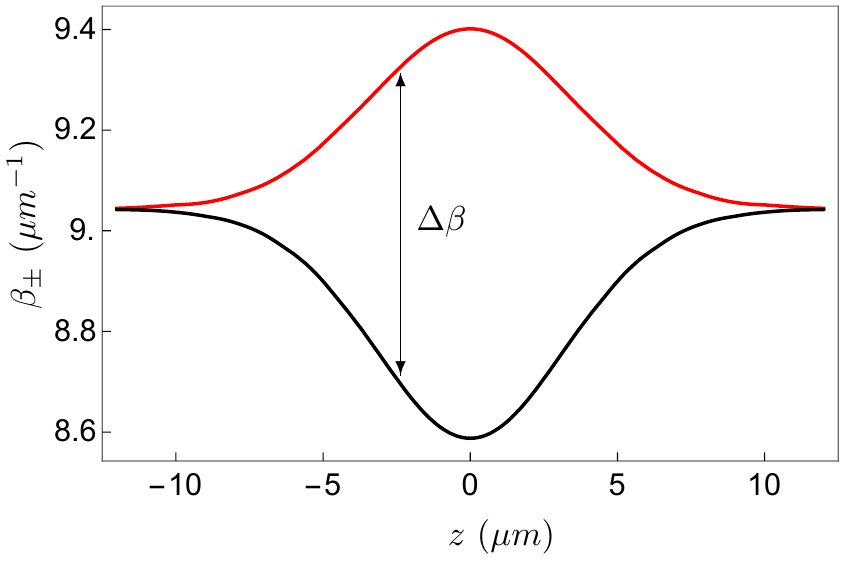}
\caption{
Propagation constants $\beta_\pm(z)$ for the symmetric and antisymmetric modes of propagations in the straight/curved waveguide system. $n_g=3.48$, $n_\text{clad}=1.45$, resonator radius $R=100 \mu$m, $\lambda=1.55\mu$m, width of both waveguides: $w=200$nm. Minimal separation $d_0=200$nm.
}
\label{fig:split}
\end{figure}

We focus here on ideal coupling, whereby the waveguide is single mode and parasitic losses are negligible~\cite{Gorodetsky-1999,Cai-2000,Spillane-2003,Pfeiffer-2017}. The direct interaction between the waveguide and the resonator usually takes place in a narrow region of space, where the resonator can be regarded as a segment of curved waveguide as in~\cite{Rowland-1993}. Let us assume for simplicity that the two waveguides have identical width $w$, propagation constant $\bar\beta$ at infinite separation, and that they have the same refractive index, see Fig.~\ref{fig:topview}. Departures from this symmetrical situation can easily be taken into account in principle (an accurate asymptotic formula of the dispersion relation for curved waveguide is given in \cite{Kozyreff-2016}). If the local distance, $d(z)$, varies slowly compared to the wavelength, then Helmholtz equation can be treated by perturbation~\cite{Snyder-1983}. In this frame,  the field is expanded in terms of the local modes of propagation as
\beq
\psi\approx a_{+}(z)\phi_{+}(x,y,z)+a_{-}(z)\phi_{-}(x,y,z).
\eeq
Above, $\phi_+(x,y,z)$ and $\phi_-(x,y,z)$ are respectively symmetric and an antisymmetric  with respect to the middle point and are normalised such that
\begin{align}
\ps{\phi_i}{\phi_j} =\iint\phi_i^*(x,y,z)\phi_j(x,y,z)\rd x\rd y&=\delta_{i,j},
& i,j&=\pm.
\end{align}
The evolution of $a_{+}(z)$ and $a_{-}(z)$ is given by~\cite{Snyder-1983}
\beq
\frac{da_\pm}{dz} \approx \ii\beta_\pm(z) a_\pm,
\label{eq:WKB}
\eeq
where $\beta_\pm(z)$ are the local propagation constants. The calculation of the local modes and their propagation constants is a 2D problem for each value of $z$. It can be solved by standard softwares such as Lumerical, COMSOL, or the spectral index method~\cite{Burke-1995}. Note that the graph of $\beta_\pm(z)$ allows one to objectively determine the extent of the coupling region: outside it, the local propagation constant are indistinguishable from their asymptotic values, see Fig.~\ref{fig:split}.

Given the amplitudes $a_\pm(z)$ associated to the symmetric and antisymmetric modes, the amplitudes in the waveguide and in the resonator can be retrieved as
\begin{align}
a_g(z)&=\left(a_-+a_+\right)/\sqrt2,
&a_r(z)&=\left(a_--a_+\right)/\sqrt2.
\end{align}
Combining the above relations, it is straightforward to derive the following matrix relation between the amplitudes at the entrance ($z=-z_c/2$) and exit ($z=z_c/2$) of the coupling zone:
\beq
\begin{pmatrix}
a_{g,2}\\a_{r,2}
\end{pmatrix}
=M\begin{pmatrix}
a_{g,1}\\a_{r,1}
\end{pmatrix}.
\label{eq:in-out}
\eeq
The matrix $M$ is given by
\beq
M=\frac{1}{2}
\begin{pmatrix}
1&1\\1&-1
\end{pmatrix}
\begin{pmatrix}
e^{\ii\int_{-z_c/2}^{z_c/2}\beta_{+}\rd z}&0\\0&e^{\ii\int_{-z_c/2}^{z_c/2}\beta_{-}\rd z}
\end{pmatrix}
\begin{pmatrix}
1&1\\1&-1
\end{pmatrix}.
\eeq
Above, $z_c$ can be any sufficiently large value that $|\Delta\beta(z_c)|\ll1$. If we write 
\beq
\beta_\pm(z)=\bar\beta\pm\Delta\beta(z)/2
\eeq
and introduce
\beq
\delta=\frac12\int_{-z_c/2}^{z_c/2}\Delta\beta(z) \rd z,
\eeq
then $M$ assumes the familiar form
\beq
M
=e^{\ii\bar\beta z_c} \begin{pmatrix}
\cos\delta&\ii\sin\delta\\
\ii\sin\delta&\cos\delta
\end{pmatrix}.
\eeq
Following Yariv~\cite{Yariv-2000}, Eq.~(\ref{eq:in-out}) is completed by the feedback condition
\beq
a_{r,1}=\tilde\alpha\; a_{r,2},
\eeq
where the complex constant $\tilde\alpha$ accounts for propagation in the resonator outside the coupling region. Thus, one easily obtains the intensity transmission coefficient of the waveguide/resonator system
\begin{align}
a_{g,2}/a_{g,1}&=e^{\ii\bar\beta z_c} \frac{\cos\delta-\tilde\alpha e^{\ii\bar\beta z_c} }{1-\tilde\alpha e^{\ii\bar\beta z_c}\cos\delta},\\
\to
T&=\left|\frac{\cos\delta-\tilde\alpha e^{\ii\bar\beta z_c} }{1-\tilde\alpha e^{\ii\bar\beta z_c}\cos\delta}\right|^2.
\label{eq:Tgeneral}
\end{align}

To make contact with notations in previous works~\cite{Yariv-2000,Cai-2000}, we write
\begin{align}
\cos\delta&=|t|e^{\ii\xi}, &\sin\delta&=\kappa,
&\text{and }& &\tilde\alpha e^{\ii\bar\beta z_c}&=|\tilde\alpha|e^{\ii\varphi},
\label{eq:cosdelta}
\end{align}
where $\xi=0$ or $\pi$, according to the sign of $\cos\delta$.
Above, $t$ is  the single-pass transmission coefficient and  $\varphi\pm\xi$ is the phase accumulated by a travelling wave in the  microresonator over a complete roundtrip, whether  it is given by $\phi_-(x)$ or $\phi_+(x)$ in the coupling region. $\varphi$ is a function of the injection frequency, $\omega$, through the dispersion relation within the resonator. Resonant injection, $\omega=\omega_R$, happens if 
\begin{align}
\varphi+\xi&=2\ell\pi,
&\ell\in\mathbb{N}.
\end{align}
The transmission at resonance is then
\beq
T(\omega_R)=\left(\frac{|\cos\delta|-|\tilde\alpha| }{1-|\tilde\alpha \cos\delta|}\right)^2.
\label{eq:Tres}
\eeq
Hence,  the condition for critical coupling, $T(\omega_R)=0$, is given by the well-known formula~\cite{Yariv-2000}
\beq
|t|=|\tilde\alpha|.
\label{eq:crit}
\eeq

Typical microcavities have very large $Q$ factor, so that $|\tilde\alpha|=\exp(-n_gk\pi R/Q)\approx1$. Consequently, critical coupling requires $|\cos\delta|\approx1$ and, in usual situations, this is achieved for a very small single-pass phase shift: $\delta\ll1$. However, it is easy to see in Eq.~(\ref{eq:Tres}) that $T(\omega_R)$ is a $\pi$-periodic function of $\delta$. If the single-pass interaction is sufficiently strong that $\delta>\pi$, then multiple critical coupling arises. 

With very good accuracy, $\Delta\beta$  decreases exponentially with $d$, at least for large enough $d$:
\beq
\Delta\beta(z)\approx  \Delta\beta_0  e^{-m d(z) },
\label{eq:Deltabeta}
\eeq
where $\Delta\beta_0$ is the splitting of propagation constants at contact. In the simplified situation where the waveguide and ring have infinite height, the decaying constant $m$ is simply given, by
\beq
m=\sqrt{\bar\beta^2-n_\text{clad}^2k^2},
\label{eq:m}
\eeq
and this expression remains applicable with reasonable accuracy even for realistic situations such as in Fig.~\ref{fig:schematic1}.  Given (\ref{eq:Deltabeta}), with $d\approx d_0+z^2/(2R)$, one immediately obtains
\beq
\delta\approx \frac12\int_{-\infty}^{\infty}\Delta\beta(z)\rd z= \Delta\beta_0 e^{-md_0} \sqrt{\frac{\pi R}{2m}}.
\label{eq:delta2}
\eeq
In this expression, $\exp(md_0)/\Delta\beta_0$ can be viewed as the effective coherence length of interaction between the waveguide and the portion of resonator with which it interacts. On the other hand, $\sqrt{\pi R/2m}$ is the effective coupling length~\cite{Boucher-2014}. Multiple critical coupling requires the effective coupling length to exceed the effective coherence length. 

\begin{figure}
\includegraphics[width=8.8cm]{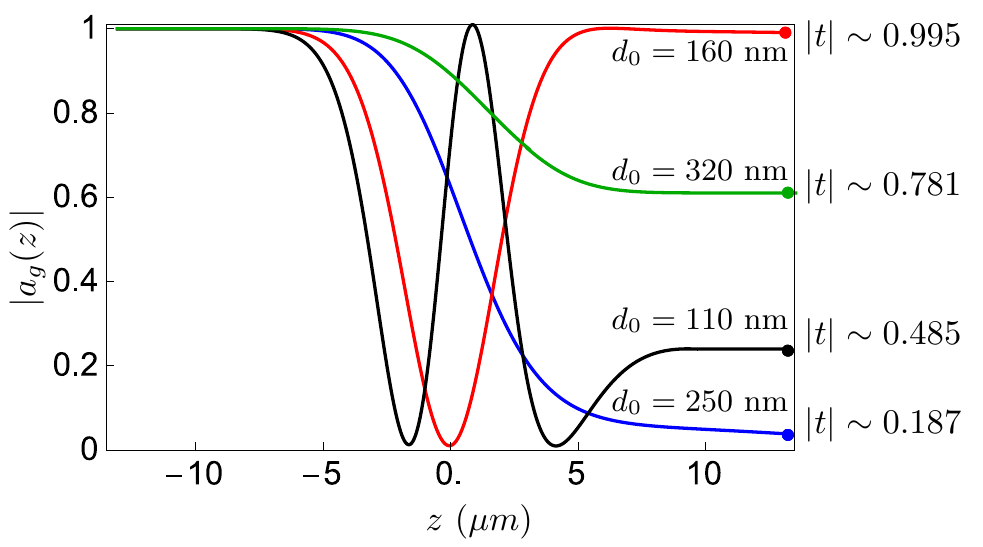}
\caption{Waveguide field amplitude in the single-pass configuration for various values of $d_0$. The single-pass transmission coefficient is given by $|t|=|\cos\delta|=\lim_{z\to\infty}|a_g(z)|$. $n_g=3.48$, $n_\text{clad}=1.45$, $R=100 \mu$m, $\lambda=1.55\mu$m, $w=200$nm, infinite height. As $d_0$ is decreased, $|t|$ varies non monotonically, allowing the critical condition $|t|=|\tilde\alpha|\leftrightarrow T(\omega_R)\approx0$ to be achieved for several values of $d_0$, see Fig.~\ref{fig:schematic1}\,(b).
}
\label{fig:singlepass}
\end{figure}

\begin{figure}
\includegraphics[width=8.8cm]{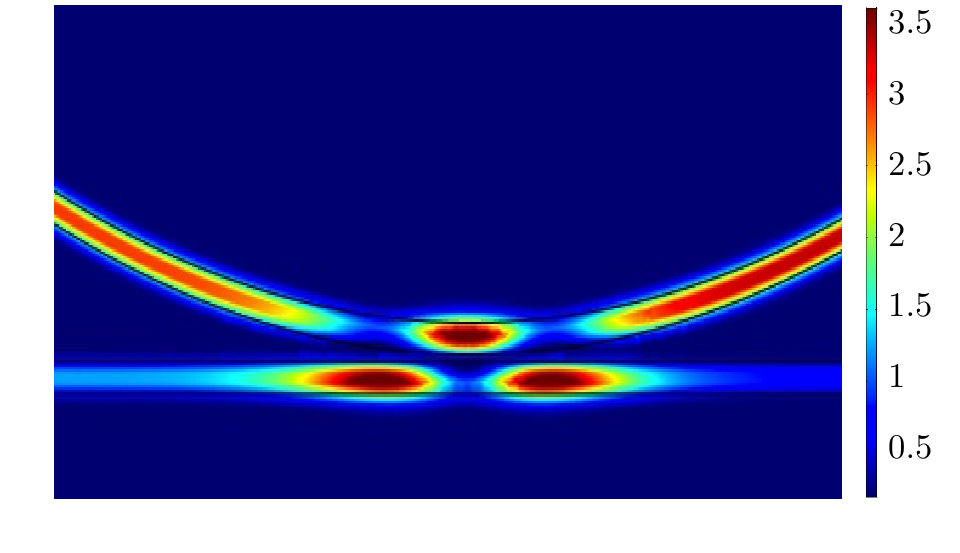}
\caption{Field intensity distribution in a waveguide resonantly coupled to a ring resonator at a high-order critical coupling distance. The higher-order character of the coupling is attested by the presence of ``hot-spots" in the waveguide. $R=30\mu m$, $d_0=46$\,nm, $w=200$\,nm and infinite height assumed.
}
\label{fig:colorful}
\end{figure}

This situation is depicted in Fig.~\ref{fig:singlepass}.  There, the field amplitude is computed in the single-pass configuration, \textit{i.e.} without the feedback provided by the cavity for various values of $d_0$. For simplicity of calculation, we assumed in that figure an infinite height, both for the waveguide and the microresonator. For sufficiently small $d_0$ the coupling length exceeds the coherence length, so that optical energy is transferred back and forth several time between the waveguide and the ring. Furthermore, Fig.~\ref{fig:colorful} show a finite element simulation (COMSOL mutliphysics 5.3) of a ring/waveguide at ``third-order" critical coupling, in the sense that $\delta\approx (3-1)\pi$, or equivalently, that the effective coupling length is twice the effective coherence length. Note the presence of three ``hot-spots"; these  result from interferences within the coupling region, combined with intensity build-up in the cavity.

From what precedes, the minimal radius for multiple critical coupling is the one for which $\delta=\pi$ as $d_0\to0$, \textit{i.e.} at contact coupling:
\begin{align}
\Delta\beta_0 \sqrt{\frac{\pi R_\text{min}}{2m}} =\pi.
\label{eq:rmin1}
\end{align}
Above, to evaluate $\Delta\beta_0$, we note that as $d_0\to 0$, the resonator plus the waveguide locally make a single waveguide of width $2w$. Then, $\Delta\beta_0$ is the separation between the fundamental even and odd modes (e.g. TE$_0$ vs TE$_1$) of that waveguide. A precise estimation requires one to solve the transcendental equation for the propagation constants of the waveguide. As a general rule, 
\beq
\Delta\beta_0\approx \frac{g}{w},
\eeq
for some constant $g$ that depends on the refractive index and geometry of the waveguide.  Substituting in Eq.~(\ref{eq:rmin1}), we finally obtain
\beq
R_\text{min} =\left(\frac{2\pi m \lambda}{g^2}\right)\times  \frac{w^2}{\lambda} \equiv  \frac{4\pi^2\sqrt{n_\text{eff}^2-n_\text{clad}^2}}{g^2} \times  \frac{w^2}{\lambda} ,
\label{eq:rmin}
\eeq
where the left factor is dimensionless, $m$ is defined in Eq.~(\ref{eq:m}) and the effective refractive index is defined as $n_\text{eff}=\bar\beta/k$. The increase of $R_\text{min}$ in $w$ is consistent with intuition, since a larger value of $w$ leads to a smaller evanescent field outside the waveguide, hence a weaker coupling and a longer coherence length. 

\begin{figure}
\includegraphics[width=8.5cm]{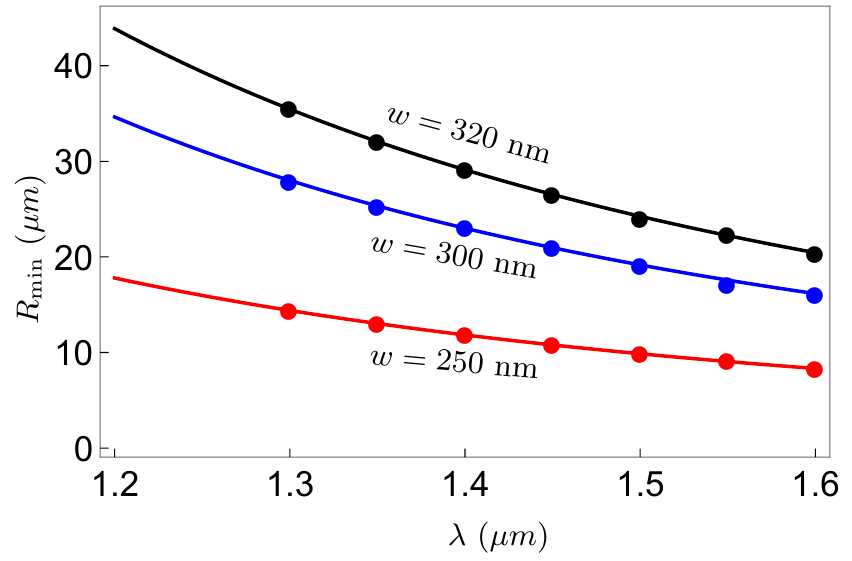}
\caption{Minimal ring radius leading to multiple critical coupling as a function of wavelength for various waveguide widths. Infinite height assumed. $n_g=3.48$, $n_\text{clad}=1.45$. Full line:  analytical results. Dots: COMSOL simulation.}
\label{fig:Rmin}
\end{figure}

The graph of $R_\text{min}$ is computed numerically for waveguides of infinite height in Fig.~\ref{fig:Rmin}, confirming the trend given in Eq.~(\ref{eq:rmin}). The values of  $\Delta\beta_0$ and $m$ were evaluated by solving the transcendental equation for a slab waveguide of width $2w$ and substituted in Eq.~(\ref{eq:rmin1}) to obtain $R_\text{min}$. Alternatively, we ran COMSOL simulations to compute the single-pass transmittance $t$ of a straight slab waveguide in contact with a curved slab waveguide. We increased the radius of curvature until obtaining $t=-1$  (\textit{i.e.} $\delta=\pi$)  and found very good agreement with the analytical results.

If we consider ridge waveguides, the transmission curve shown in Fig.~\ref{fig:schematic1}(b) can accurately be reproduced by substituting Eq.~(\ref{eq:delta2}) into Eq.~(\ref{eq:Tres})  with appropriately chosen values of $\Delta\beta_0$ and $m$. For the specific parameters of Fig.~\ref{fig:schematic1}, fitting values are $\Delta\beta_0=4.47\mu m^{-1}$ and $m=8.42\mu m^{-1}$.

%---------------------------------------
\section{Sensing and switching  applications}
\begin{figure}
\includegraphics[width=8.5cm]{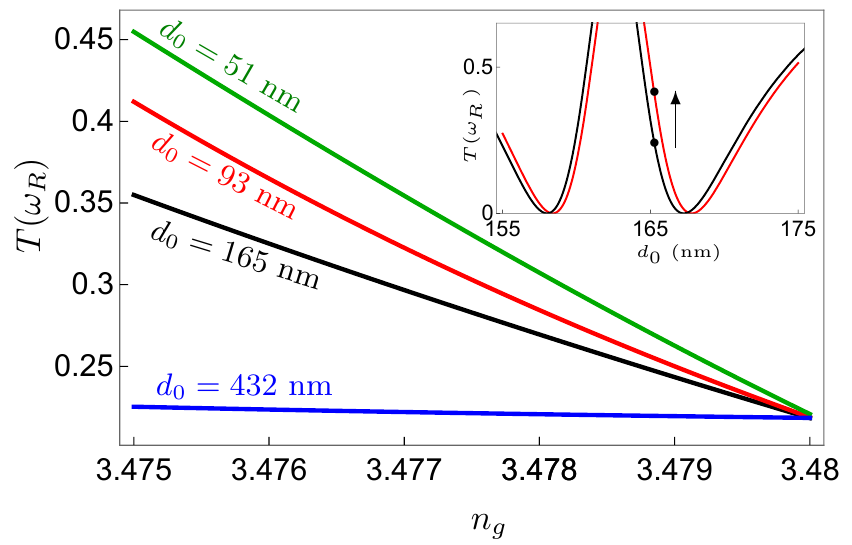}
\caption{Resonant transmission as a function of $n_g$ for values of $d_0$ near critical values. Inset: transmission curve in the vicinity of $d_0=165$\,nm, $w=200$\,nm, $R=100\mu m$, $n_clas=1.45$, $\lambda\approx1.55nm$ and $n_g=3.48$ (black) or $n_g=3.475$ (red curve).}
\label{fig:sensing}
\end{figure}
Once the conditions for multiple critical coupling configuration are established, we note that the sharp features in the dependence of $T$ on $d_0$, (see Fig.~\ref{fig:schematic1} and inset of Fig.~\ref{fig:sensing}), could serve as the basis of a new detection principle. Indeed, a change in the cladding refractive index $n_\text{clad}$, or in the guiding index $n_g$ would amount to effectively change $d_0$. 

Rather than to monitor the spectral shift of resonances in the transmission spectrum, we propose to monitor changes in the transmission dip at resonance. This is slightly different from previous intensity-detection scheme, such as Ref.~\cite{Chao-2006}, where one monitors the transmitted intensity at a fixed, near-resonant, wavelength, as the cladding index is varied. In that case, variations of transmitted intensity are due to resonance shift. Here, we propose to follow the resonance peak and monitor the depth of the resonant transmission dip as the refractive index is changed.

Let the incident power be centred on the resonance and given by $I(\omega_R)$. A change of the guiding index
\beq
n_g\to n_g+\Delta n_g
\eeq
leads to a transmission change
\beq
T(\omega_R)\to T(\omega_R)+\pd{T(\omega_R)}{n_g}\Delta n_g,
\eeq
and, hence, to change in transmitted power
\beq
I(\omega_R)\pd{T(\omega_R)}{n_g}\Delta n_g
\eeq
Given the Noise Equivalent Power (NEP) of the photodetector and the measurement bandwidth $\Delta f$, the smallest detectable index change is given by~\cite{Mackowiak-2015}
\beq
\Delta n_{g,\text{min}} =\frac{ \text{NEP}\times\sqrt{\Delta f} }{ I(\omega_R) \left|\pd{T(\omega_R)}{n_g}\right|}
\eeq
While the value of $\left|\dd T/\dd n_g\right|$ is rather modest around the first critical distance, it raises sharply in the vicinity of higher order critical coupling distances, see Fig.~\ref{fig:sensing}. Note from the inset of Fig.~\ref{fig:sensing} that the slope $|\dd T/\dd d_0|$ is highest on the under-coupling side of the critical point.

Assuming an NEP on the order of $10\text{pW}/\sqrt{\text{Hz}}$~\cite{Mackowiak-2015}, an input source power of $I(\omega_R)=10$mW with a frequency bandwidth of 1GHz, Table~\ref{table1} gives Limits Of Detection (LOD) $\Delta n_\text{g,min}$ in the vicinity of various critical coupling distances. With the numbers assumed here ($\lambda\approx 1.55\mu m$, cavity radius $R=100\mu m$, $n_g=3.48$, waveguide width $w=200$\,nm) one finds an LOD of $7.5\times10^{-7}$ RIU. This should not be regarded as an ultimate value, as better sources can in principle be used, with higher incident power and narrower bandwidth. Also, we see from Table~\ref{table1} that the LOD improves substantially with the order of critical coupling. Higher-order critical coupling, corresponding to smaller $d_0$ are liable to yield even smaller LOD. However, the rapidity of oscillation becomes such that numerical investigations become challenging in that region of parameters.

\begin{table}
\caption{\label{table1} Limit of detection through intensity measurements in the vicinity of critical points. Same resonator and waveguide parameters as in Fig.~\ref{fig:singlepass}. $\text{NEP}=10\text{pW}/\sqrt{\text{Hz}}$. Detection bandwidth $\Delta f=1$GHz. Input power $I(\omega_R)=10$mW } 
\begin{ruledtabular}
\begin{tabular}{cccc}
Critical region & $d_0$ (nm) & $\left|\pd {T}{ n_g}\right|$ & $\Delta n_{g,\text{min}}$\\ 
\hline
1 &  432 & 1 & $3.2\times 10^{-5}$\\
2 & 165 & 25 &$1.3\times 10^{-6}$ \\
3 & 93 & 31 &$1.\times 10^{-6}$\\
4 & 51 & 42 &$7.5\times 10^{-7}$ \\
\end{tabular}
\end{ruledtabular}
\end{table}

Alternatively to the above application, one may envisage to actively induce a change of refractive index in order to induce a desired change $\Delta I_\text{out}(\omega_R)$ in the output intensity. In this switching set-up, the required change is, simply
\beq
\Delta n_{g} =\frac{ \Delta I_\text{out}(\omega_R) }{ I(\omega_R) \left|\pd{T(\omega_R)}{n_g}\right|}.
\eeq

It is a simple matter to show that $|\dd T/\dd d_0|$ or $|\dd T/\dd n_g|$ scales as $\sqrt Q$: In the region of higher-order coupling, critical coupling configurations come in pairs, flanking a state of complete transmission $T(\omega_R)=1$, see Fig.~\ref{fig:schematic1}. Let us assume that, for the appropriate value of $d_0$, the value $n_g^*$ of the guiding refractive index makes $|t|=1$, and, hence, $T=1$. In the vicinity of this value, we have
\beq
|t|\sim1-|t''(n_g^*)|\frac{\left(n_g-n_g^*\right)^2}{2}.
\eeq
On the other hand,
\begin{align}
|\tilde\alpha|&\sim1-\eps,& \text{with}&&\eps\propto1/Q\ll1.
\end{align}
The nearest critical coupling configuration, $|t|=|\tilde\alpha|$, thus happens for
\beq
|n_g-n_g^*|\sim \sqrt{2\eps/|t''(n_g^*)|}
\eeq
Since the resonant transmittance changes from 1 to 0 over this change of refractive index, the average slope of this dependence is proportional to $1/\sqrt{\eps}$, \emph{i.e.} to $\sqrt{Q}$.

%----------------------------------------
\section{Conclusions}
In this work, we have expanded the classical theory of waveguide/resonator coupling and showed that, contrary to common assumption, multiple critical coupling distances \emph{can} exist when the bus waveguide lies in the same plane as the resonator. This effect exists as soon as the resonator radius exceeds a critical radius, for which we give an analytical estimate as a function of wavelength and waveguide transverse dimension. The present treatment, being expressed  in terms of the local splitting of propagation constants, $\Delta\beta(z)$,  can directly be   transposed to other geometries, \textit{e.g.}  racetrack resonators or  waveguide and ring lying in different planes. In the case of a circular resonator side coupled to a waveguide, the function $\Delta\beta(z)$ has a gaussian profile with width controlled by the cavity radius. Thus, the effective coupling lengths was found to be $\sqrt{\pi R/2m}$.  For racetrack resonators, the splitting function $\Delta\beta(z)$ exhibits a plateau around $z=0$. In all cases, the space-dependence of $\Delta\beta(z)$ make the single-pass coupling problem distinct from that of coupled straight parallel waveguides.

As we have shown, multiple critical coupling can be exploited as a new detection principle. Equally, the on- or off-switching of resonances in the transmission spectrum could be used as  an optical gate. More generally, this study shows that the single-pass transmission parameter $t$, and hence the coupling parameter $\kappa=\sqrt{1-t^2}$, may vary in a much more complicated way than anticipated as a function of the waveguide/resonator  distance $d_0$. This behaviour may become important to consider in future designs of photonic integrated circuits in which micro-resonators are expected to play major roles. Although we have illustrated it with silicon refractive index and telecom wavelength, the theory presented here is general and independent of the material considered. Let us note that water strongly absorbs light at $\lambda=1.55\mu m$, so that sensing in aqueous environment is more suitably done at $\lambda=1\mu m$. We have therefore checked that all our conclusions hold with  Al$_2$O$_3$ waveguides with realistic fabrication parameters operating around 1$\mu m$. \\ 
\mbox{}
\\

%%%%%%%%%%%%%%%%%%%%%%%%%%%%%%%%%%%%%%%%%%%%%%
\section*{Acknowledgements}

This project has received funding from the European Union's Horizon 2020 research and innovation programme under grant agreement No 634928. GK is a research associate with the Fund for Scientific Research-FNRS. We thank Pascal Kockaert for useful discussions, as well as Johann Toudert, Johann Osmond, Michiel de Goedde and Sonia Garc\' ia Blanco for communicating precious details on device fabrication.

%\appendix
%\section{dispersion relation of a slab waveguide}
%{\color{red}
%The dispersion relation of a slab waveguide of thickness $2w$, refractive index $n_g$ embedded in an infinite medium of refractive index $n_\text{clad}$ is given by the transcendental equation
%\beq
%\tan(\kappa w)= \xi m / \kappa,
%\label{slabdisp}
%\eeq
%with $\kappa^2=n_g^2 k^2-\beta^2$, $m^2=\beta^2- n_\text{clad}^2k^2$. For TE modes, $\xi=1$; for TM modes, $\xi=n_g^2/n_\text{clad}^2$. 
%}

%\bibliography{../../GLAM}


\end{document}